\documentclass[twocolumn]{aastex631}
\draft
\usepackage{natbib}
\usepackage{graphicx}
\usepackage{url}
\usepackage{xcolor}
\newcommand{\cmt}{\,cm$^{-3}$}   
\newcommand{\cmd}{\,cm$^{-2}$}   

\newcommand{\kms}{\,km\,s$^{-1}$}

\newcommand{\myr}{\,$M_{\odot}\,{\rm yr}^{-1}$}
\newcommand{\es}{$\rm\,erg\,s^{-1}$}

\newcommand{\ha}{H$\alpha$}
\newcommand{\hb}{H$\beta$}

\newcommand{\hi}{H\,{\scriptsize I}}
\newcommand{\hei}{He\,{\scriptsize I}}
\newcommand{\heii}{He\,{\scriptsize II}}
\newcommand{\civ}{C\,{\scriptsize IV}}
  
\newcommand{\oiii}{O\,{\scriptsize III}}
\newcommand{\oi}{O\,{\scriptsize I}}
\newcommand{\ro}{\,$R_{\odot}$}
\newcommand{\mo}{\,$M_{\odot}$}

\shorttitle{Z~And-type outburst in the classical nova V1047~Cen}
\shortauthors{Skopal and Shagatova}
\begin{document} 

\title{V1047~Cen: The first Z~And-type outburst observed 
       in the classical nova binary}
\author[0000-0002-8312-3326]{Augustin Skopal}
\affiliation{Astronomical Institute, Slovak Academy of Sciences, \\
             059\,60 Tatransk\'a Lomnica, Slovakia}

\author[0000-0002-1218-8296]{Natalia Shagatova}
\affiliation{Astronomical Institute, Slovak Academy of Sciences, \\
             059\,60 Tatransk\'a Lomnica, Slovakia}
\begin{abstract}
In 2019, the classical nova V1047~Cen experienced an unusual 
outburst, the nature of which has not yet been clearly determined. 
In this paper, we show that the 2019 V1047~Cen outburst is of 
Z~And-type -- a type that is characteristic and has so far been 
observed only in symbiotic binaries. 
% Methods + Results
We support our claim by modeling the near-ultraviolet 
to near-infrared spectral energy distribution, which revealed 
a close similarity between the fundamental parameters and the 
mass-loss rate of the burning white dwarf during the 2019 V1047~Cen 
outburst and those measured during Z~And-type outbursts in 
symbiotic stars. All parameters are in good agreement with 
the theoretical prediction when the accretion rate exceeds 
the stable burning limit for white dwarfs with masses 
$\lesssim 0.7$\mo. 
% Conclusions
Our analysis showed that after a nova explosion, the Z~And-type 
outburst can occur not only in symbiotic binaries but also in 
short-period cataclysmic variables, when the accretion-powered 
system changes to a nuclear-powered, as a consequence of 
the donor's reaction to the nova explosion. Such a development 
promotes the production of Type Ia supernovae. 
\end{abstract}

\section{Introduction}
\label{s:intro}
\subsection{Outbursts in accreting white dwarf binaries}
\label{ss:awdbs}
Non-magnetic accreting white dwarf binaries in which the white 
dwarf (WD) accretes from a main-sequence star in Cataclysmic 
Variables (CVs) or from an evolved giant in Symbiotic Stars 
(SySts) can produce several types of outbursts 
\citep[see][ for reviewes]{1995cvs..book.....W,2017PASP..129f2001M,
                           2020AdSpR..66.1097B,2023hxga.book..129B,
                           2023awdf.book.....S}. 
The response of the WD to mass accretion has been investigated by 
many authors. It was found that the outcome of accretion and its 
timescales are basically determined by the accretion rate onto 
the WD, $\dot{M}_{\rm acc}$, its mass, $M_{\rm WD}$, and the WD 
core temperature 
\cite[e.g.,][]{1978ApJ...222..604P,1980A&A....85..295S,
               1982ApJ...253..798N,2005ApJ...623..398Y,
               2007ApJ...663.1269N,2007ApJ...660.1444S}. 
Accordingly, for this study, we first summarize the main physical 
phenomena that result from accretion onto a WD of different mass 
in CVs and/or SySts as follows. 
%

% DNe:
{\sf Dwarf novae.} When $\dot{M}_{\rm acc}$ is as low 
as $\lesssim 10^{-10}$\myr, smaller than the mass transfer rate 
from the donor, the accretion disk gradually increases its mass 
up to a critical value at which it becomes hot and luminous due 
to a viscosity-induced instability, allowing a temporal increase 
in accretion onto the WD on a timescale of days to weeks. 
We observe optical brightening in a range of typically 2--6\,mag, 
the so-called dwarf nova (DN) outbursts. 
These accretion-powered events represent the most frequently 
observed outbursts in CVs, recurring every few weeks to months, 
and generating the luminosity of $\approx 10^{34}$\es\ 
\citep[e.g.,][]{1974PASJ...26..429O,2001NewAR..45..449L,
                2004ApJ...610..422M,2012ApJ...747..117C}. 

% Classical novae (CNe):
{\sf Classical novae.} 
Prolonged accretion at a wide range of rates (say, from a few 
times 10$^{-13}$ to a few times 10$^{-7}$\myr) gradually 
increases the pressure and temperature of the layer of matter 
accumulated on the WD surface up to critical values of 
$\sim$10$^{19}$\,dyn\cmd\ and $\sim$10$^8$\,K, at which 
thermonuclear runaway (TNR) is triggered at its base, causing 
a brightening of the system by up to $\sim$15\,mag. We observe 
the so-called classical nova (CN) explosion 
\citep[e.g.,][]{1972ApJ...176..169S,
                1978ARA&A..16..171G,
                1989PASP..101....5S,
                2008clno.book.....B,
                2016PASP..128e1001S}. 
Theoretical modeling of \cite{1995ApJ...445..789P} and 
\cite{2005ApJ...623..398Y} confirmed that the temporal evolution 
(slow, fast, and recurrent novae, including their extremes), as 
well as the main properties (e.g., the peak luminosity, the rate 
of the ejected matter and its expansion velocity) of CNe, can be 
reproduced by suitable combinations of three independent 
parameters: $\dot{M}_{\rm acc}$, $M_{\rm WD}$, and the temperature 
of the WD core. 

% --- slow CNe
Accretion onto a low-mass WD ($\lesssim$1\mo) at rather high 
rates (say, $\gtrsim$10$^{-9}$\myr) produces slow CNe (the time 
to decline by 2\,mag from maximum, $t_2 > 80$ days) with 
a smaller amplitude ($\lesssim$10\,mag). 
This is because the bottom of the accreted layer is less degenerate 
due to less surface gravity, and the temperature may not even reach 
$10^8$\,K when (non-explosive) TNR occurs, generating little 
nucleosynthesis 
\citep[][]{1982ApJ...257..752F,
           1982ApJ...257..767F,
           2016PASP..128e1001S}. 
Several extreme cases, with a rise to visual maximum over 
months and a decline to quiescence lasting decades to centuries, 
were identified as SySts \citep[see][]{1980MNRAS.192..521A}, 
and are therefore called symbiotic novae (SyNe). 
%---
Following studies showed that TNR can also power SyNe 
\citep[e.g.,][]{1983ApJ...273..280K}, having physical parameters 
as are typical of CNe in CVs, including the slowest SyN AG~Peg 
\citep[1850 to $\sim$1985, see][]{1993AJ....106.1573K}. 
However, the finding that the total radiation output of SyNe 
($10^{46.5} - 10^{47}$\,erg) exceeds that of CNe in CVs 
\citep[see][]{1994A&A...282..586M} may be the result of 
efficient mass-transfer via the wind from the evolved giant 
due to its focusing towards the orbital plane 
\citep[e.g.,][]{2012BaltA..21...88M,2016A&A...588A..83S}, 
which can support the residual burning on the WD surface and 
thus prolonging the nova lifetime. The latest overview of SyNe was 
presented by \cite{2024arXiv241220499M}. 
%

% --- fast CNe on massive WDs :
In contrast, accretion onto a more massive WD ($>$1\mo) at lower 
rates (say, $\lesssim$10$^{-9}$\myr) ignites an explosive TNR 
under degenerate conditions, producing fast CNe ($t_2 < 25$ days) 
with a higher amplitude ($\gtrsim$10\,mag). If the accretion 
occurs at a high rate, a few times (10$^{-8} - 10^{-7}$)\myr, 
the TNR repeats on a timescale of $<$100 years. These events 
are called recurrent (symbiotic) novae 
\citep[e.g.,][]{2005ApJ...623..398Y}. 

% SSS :
{\sf Supersoft X-ray sources.} 
For specific cases of high $\dot{M}_{\rm acc}$ between a few times 
10$^{-8}$ and a few times 10$^{-7}$\myr\ and $M_{\rm WD}$ between 
$\sim$0.6 and $\sim$1.35\mo, respectively, hydrogen-rich material 
can burn stably on the WD surface, i.e., at the same rate as it 
is accreted \cite[see e.g., Fig.~1 of][]{2013ApJ...777..136W}. 
The corresponding nuclear energy output, typically of 
$10^{36}-10^{38}$\es, is emitted in the supersoft X-ray energies 
below $\sim$1\,keV. Therefore, these objects are called 
supersoft X-ray sources 
\citep[SSSs, e.g.,][]{1992A&A...262...97V,1996LNP...472.....G}. 
%---
Due to extreme absorption in this domain, not all stable burning 
WDs can be observed as SSSs. %, which is limited by their location. 
This is the case for many nuclear-powered SySts, of which only 
a few are reliably observed as SSSs. For example, symbiotic X-ray 
binaries, AG~Dra 
\citep[$N_{\rm H} = 3.15\times10^{20}$\cmd,][]{2008A&A...481..725G}, 
LIN~358 
\citep[$N_{\rm H} = 6.1\times10^{20}$\cmd,][]{2015NewA...36..116S}, 
and Draco~C1 
\citep[$N_{\rm H} = 4.3\times10^{20}$\cmd,][]{2018MNRAS.473..440S}. 
% ---
Basically, we observe persistent SSSs % In principle, we observe ...
\citep[see a catalog of][]{1996LNP...472..299G} 
or transient (recurrent) sources, such as the SSS phase of CNe. 

% - Z And type outbursts:
{\sf Z~And-type outbursts.} 
When $\dot{M}_{\rm acc}$ exceeds the level sustaining stable burning 
($\gtrsim 10^{-7}$)\myr, the material will still burn stably 
(i.e., generate the luminosity of a few times $10^{37}$\es), but 
not at the same rate as it is accreting. The burning cannot 
consume all the accreted material, which leads to the formation 
of a red-giant-like structure 
\citep[e.g.,][]{1980A&A....82..349P,
               1982ApJ...253..798N} 
or the blowing of an optically thick wind from the WD at rates 
$\gtrsim 10^{-6}$\myr\ 
\citep[e.g.][]{1994ApJ...437..802K,
               1996ApJ...470L..97H}\footnote{The principle of 
Z~And-type outbursts was already 
postulated by \cite{1976Afz....12..521T}. However, the lack of 
knowledge of the accurate WD's parameters during these outbursts 
(luminosity and mass-loss rate) meant that their true nature was 
not clear for a long time (see Sect.~4.3.2 of 
\cite{1991A&A...248..458M}, Sect.~3.3 of \cite{2016MNRAS.462.4435T}, 
and Sect.~4.2 of \cite{2017A&A...604A..48S}).}. 
These events are observationally indicated by a few magnitudes 
brightening in the optical, evolving on a timescale of weeks 
to years 
\citep[e.g.,][]{1991BCrAO..83..104B,
                2008JAVSO..36....9S,
                2008MNRAS.385..445L,
                2019CoSka..49...19S}, 
always accompanied by an enhanced mass outflow 
\citep[e.g.,][]{1995ApJ...442..366F,
                2011PASP..123.1062M} 
at rates of a few times $10^{-6}$\myr, and terminal 
velocities of $(1 - 3)\times 10^3$\kms\ 
\citep[][]{2006A&A...457.1003S,2023MNRAS.526.6381S}. 
They are called Z~And (-type) outbursts because they were observed 
in the past for the prototype of the class of SySts -- Z~And 
\citep[see Fig.~1.2 of][]{1986syst.book.....K}. %Z~And-type outbursts 
So far, they have been observed exclusively in the light 
curves of SySts. 
%They occur on an irregular timescale -- in some 
%nuclear-powered systems have not yet been observed 
%\citep[e.g., SY~Mus, RW~Hya,][]{1995A&A...293..783P, 
%                                1995AJ....110..391K}. 
% 
It was found that during Z~And-type outbursts the luminosity of 
the burning WD, $L_{\rm WD}$, is several times 10$^{37}$\es, its 
effective radius, $R_{\rm WD}^{\rm eff}$, is around 0.1\ro, 
the blackbody temperature, $T_{\rm WD}$, is $(1-2)\times 10^5$\,K, 
and the emission measure of the nebular continuum, $EM$, is 
a few times $10^{60}$\cmt\ 
\citep[][]{1991A&A...248..458M,
           2005A&A...440..995S,
           2012ApJ...750....5K,
           2017A&A...604A..48S,
           2018MNRAS.479.2728C}. 
The nebular emission represents a part of the hot WD radiation 
converted by the fast wind flowing from the WD at rates, 
${\dot{M}}_{\rm WD} \gtrsim 10^{-6}$\myr\ 
\citep[][]{1992ApJ...387..624S,
           2006A&A...457.1003S,
           2017A&A...604A..48S,
           2020A&A...636A..77S}. 
Both the $L_{\rm WD}$ and ${\dot{M}}_{\rm WD}$ derived from 
observations are consistent with the theoretical prediction 
mentioned at the beginning of this paragraph. 

Finally, we note that the post-outburst evolution of CN may be 
affected by a significant increase in mass transfer due to 
irradiation of the donor by the nova explosion 
\citep[][]{1988ApJ...325..828K,
           2020NatAs...4..886H,
           2021MNRAS.507..475G}. 
This effect can explain the transition from an accretion-powered 
system before the SyN outburst to a nuclear-powered system after 
it, which may subsequently lead to the triggering of a Z~And-type 
outburst 
\citep[see Sect.~4.9 of][]{2017A&A...604A..48S}. 
In the case of some CNe 
\citep[e.g., GQ~Mus and V723~Cas,][]{1993Natur.361..331O,
                                     2008AJ....135.1328N} 
and very luminous SSSs, this effect may support residual surface 
nuclear burning, which helps to understand their unusually 
long-lasting and luminous SSS phases 
\citep[see][]{2022AJ....164..145S}. 

Here, in the case of the CN V1047~Cen, we show that its 
unusual 2019 outburst is analogous to the Z~And-type outburst 
in SySts that occur after their SyN outburst 
\citep[see Fig.~12 of][]{2020A&A...636A..77S}. 

\subsection{Classical nova V1047~Cen and its 2019 outburst}
\label{ss:v1047}
The explosion of the CN V1047~Cen (Nova Cen 2005) was discovered 
by \cite{2005CBET..215....1L} on 2005 September 1.03 at a visual 
magnitude of 8.5. The authors supported the nature of the transient 
as a CN with a spectrum obtained a few days later. Subsequent 
spectroscopic observations specified the `Fe\,{\scriptsize II}' 
type of this nova \citep{2012PASP..124.1057W}. 
According to \cite{2022ApJ...939....6A}, the progenitor's 
brightness of $V>20.5 - 21$ corresponds to the nova amplitude of 
$\Delta{V}\approx 13$\,mag and an absolute magnitude of $M_{\rm V}>5$, 
both of which favor a typical cataclysmic variable containing 
a main-sequence donor star. 

Another outburst of V1047~Cen was recorded on 2019 June 11 by 
\cite{2019TNSTR.985....1D} as a transient AT2019hik/Gaia19cfn. 
Based on the Optical Gravitational Lensing 
Experiment survey \citep[OGLE;][]{2015AcA....65....1U}, 
\cite{2019ATel12876....1M} unambiguously associated the transient 
with V1047~Cen and found that the re-brightening started already 
on 2019 April 6.11 with a gradual increase by $\sim$2.5\,mag 
measured on 2019 June 12.04 in the OGLE $I$ band. The authors noted 
that such slow brightening is not consistent with a recurrent nova 
explosion. 

This unexpected outburst led to subsequent spectroscopic monitoring 
of V1047~Cen in the optical and near-infrared (NIR) 
\citep[see][]{2019ATel12885....1A,
              2019ATel12975....1A,
              2019ATel12893....1D,
              2019ATel12923....1G}, 
while the $BVR_{\rm C}I_{\rm C}$ photometry was performed 
by amateur astronomers and collected in the $AAVSO$ 
database. %\footnote{see https://www.aavso.org/LCGv2/}. 
Analyzing NIR spectroscopy, broadband optical, and NIR photometry 
during the first 160 days of the 2019 outburst, 
\cite{2019ApJ...886L..14G} concluded that the outburst is most 
likely a dwarf nova eruption. However, the authors discussed its 
long duration and too-early appearance after the CN outburst 
as the main shortcomings of this interpretation. 
Based on multiwavelengths, near-ultraviolet to radio, observations 
spanning the entire, more than 400-day lasting outburst, 
\cite{2022ApJ...939....6A} suggested {\it "that the outburst may 
have started with a brightening of the disk due to enhanced mass 
transfer or disk instability, possibly leading to enhanced nuclear 
shell burning on the white dwarf, which was already experiencing 
some level of quasi-steady shell burning."} 
The authors pointed to observational similarities with outbursts 
in SySts. The exceptionality of this outburst led the authors to 
conclude that {\it "the 2019 outburst of V1047~Cen appears to be 
unique, and nothing similar has been observed in a typical 
cataclysmic variable system before, hinting at a potentially 
new astrophysical phenomenon."} 
%
%
%===============================================|
%-- Table 1: Parameters of the 2019 outburst ---|
%===============================================|
%
\begin{table*}[th!]
\caption{Physical Parameters of the 2019 Outburst of the Nova V1047~Cen 
Determined by SED Modeling$^{\rm a}$. 
% (Sects.~\ref{ss:param} and \ref{ss:mdot}). 
        }
\begin{center}
\begin{tabular}{ccccccc}
\hline
\hline
\noalign{\smallskip}
Age$^{\rm b}$                                 & 
$T_{\rm e}$                                   & 
$EM$                                          & 
$T_{\rm WD}$                                  & 
$R_{\rm WD}^{\rm eff}$                        & 
$L_{\rm WD}$                                  & 
${\dot{M}}_{\rm WD}$                          \\
%$R_{\rm WD}^{\rm eff}$~$^{\rm c}$             &
%$L_{\rm WD}~^{\rm c}$                         \\
%-----------------------------------------------
(JD-JD$_0$)                                   &
(kK)                                          &
($10^{60}\,{\rm cm^{-3}}$)                    & 
(kK)                                          & 
($R_{\odot}$)                                 & 
($10^{37}{\rm erg\,s^{-1}}$)                  & 
($10^{-6}M_{\odot}\,{\rm yr}^{-1}$)           \\
%($R_{\odot}$)                                 &
%($10^{37}{\rm erg\,s^{-1}}$)                  \\
%
% Age,  JD,  Te,  EM,  T_BB,  R_WD,  L_WD,  dotM,  R_WD,  L_WD 
%
\noalign{\smallskip}
\hline
\noalign{\smallskip}
~~86 & 30$\pm$5    & 1.4$\pm$0.1   & 175$\pm$10$^{\rm c}$    & 
     0.13$\pm$0.01 & 5.50$\pm$0.80 & 1.34$\pm$0.11 \\
%    0.06$\pm$0.01 & 1.04$\pm$0.12 \\
 118 & 30$\pm$5    & 3.3$\pm 0.2$  & 176$\pm$10$^{\rm c}$    & 
     0.13$\pm$0.01 & 5.45$\pm$0.90 & 2.05$\pm$0.15 \\
%    0.09$\pm$0.01 & 2.46$\pm$0.26 \\
 308 & 20$\pm$3    & 4.9$\pm 0.3$  & 136$\pm$10$^{\rm c}$    & 
     0.20$\pm$0.02 & 4.91$\pm$0.90 & 3.16$\pm$0.27 \\
%    0.19$\pm$0.02 & 4.37$\pm$0.47 \\
 407 & 40$\pm$5    & 0.13$\pm$0.02 & $\sim$190     & 
       $\sim$0.07  & $\sim$2       & $\sim$0.28    \\
%      --          & --            \\
\noalign{\smallskip}
\hline
\end{tabular}
\end{center}
{\bf Notes.} \\
$^{\rm a}$ Designation as in Sections~\ref{ss:param} and \ref{ss:mdot}, \\
$^{\rm b}$ JD$_0$ = 2\,458\,579.61 (2019 April 6.11) is 
           the date of the outburst beginning. \\
$^{\rm c}$ Interpolated values from Table~\ref{tab:EMpar} to dates 
           of the SED models (full circles in Fig.~\ref{fig:lrtw}\,d). \\
\label{tab:SEDpar}
\end{table*}

Here, analyzing the observations of \cite{2019ApJ...886L..14G} and 
\cite{2022ApJ...939....6A} using the method of multiwavelength 
modeling of the spectral energy distribution (SED) in the composite 
spectra of SySts \citep[see][]{2005A&A...440..995S}, we classify 
the 2019 V1047~Cen outburst as a Z~And-type, commonly produced 
by WDs in SySts. 
To achieve our goal, we show that the fundamental parameters of 
the burning WD ($L_{\rm WD}$, $R_{\rm WD}^{\rm eff}$, $T_{\rm WD}$) 
and its ${\dot{M}}_{\rm WD}$ during the 2019 V1047~Cen outburst, 
are comparable to those of the Z~And-type outbursts, and the 
pre- and post-outburst luminosities are consistent with stable 
hydrogen burning on the WD surface. 

Accordingly, in Sect.~\ref{s:results} we describe and present the 
results of our analysis, while a discussion and conclusion are 
found in Sects.~\ref{s:discuss} and \ref{s:concl}, respectively. 

\section{Analyzis and results}
\label{s:results}
\subsection{Selection of days for SED modeling}
\label{ss:dataforSED}
Based on observations covering the 2019 outburst of V1047~Cen 
\citep[see][]{2019ApJ...886L..14G,
               2022ApJ...939....6A}, 
we modeled the SED from the near-ultraviolet (NUV) to the NIR, 
from 0.1928 to 4.60\,$\mu$m. We used {\it Swift}-UVOT $UVW2$, 
$UVM2$, optical $BVR_{\rm C}I_{\rm C}$, NIR $JHK$ and NEOWISE 
$W1$, $W2$ photometry complemented with optical 
($\sim$0.4 -- 0.9\,$\mu$m) and NIR (0.9 -- 2.5\,$\mu$m) 
spectroscopy. 

% selection of days 86, 118 and 308
Before the outburst maximum (see Fig.~\ref{fig:lrtw}\,a, dotted 
vertical line), we reconstructed the observed SED for days 86 
and 118\footnote{Days since the outburst onset, 2019 April 6.11 
(see Sect~\ref{ss:v1047}).}, 
%%JD\,2\,458\,579.61 \citep[][]{2019ATel12876....1M}.}, 
when the absolute fluxes of simultaneous NIR photometry and NIR 
spectroscopy were available and in good agreement. 
The corresponding UVOT magnitudes were estimated by interpolating 
the closest observations to these days. 
   During the maximum brightness (days 264 to 331), %Fig.~\ref{fig:colors}), 
we selected day 308 because of the simultaneous 
NUV and $W1$, $W2$ NIR photometry were available. 
Finally, after the outburst, we reconstructed the model SED using 
the $UVW2$ magnitude (day 413), emission-line-free optical 
magnitudes (days 402 to 436) and $W1$, $W2$ fluxes (day 469). 
%The optical spectrum from day 407 was scaled to the optical 
%photometry. 
%

% general means of the treatments
Optical and NIR magnitudes were converted to fluxes according to 
the calibration of \cite{1982asph.book.....H}, %and \cite{1979PASP...91..589B}, 
while the NUV magnitudes were calibrated using the UVOT photometric 
system \citep[see][]{2008MNRAS.383..627P}. 
To obtain photometric flux points of the true continuum, we 
determined corrections for emission lines from the used spectra 
(see Appendix~\ref{app:A}, Table~\ref{tab:dml}). 
The emission-line-free $BVR_{\rm C}I_{\rm C}$ photometry was 
used to verify the flux calibration of the optical spectra 
and for the color-diagram diagnostic in Sect.~\ref{ss:color}. 
In all cases, observations obtained not simultaneously with the 
selected day were calibrated using optical magnitudes corrected 
for emission lines. This is justified by the dominance of the 
nebular continuum in the optical/NIR (see Sect.~\ref{ss:param}). 

Observations were dereddened with $E_{\rm B-V}$ = 1.0 
%using the extinction curve of \cite{} Cardelli et al. (1989) 
and the distance-dependent parameters were scaled to a distance 
of 3.2\,kpc \citep[][]{2022ApJ...939....6A}. 
%
%
%==========================|
%-- Fig. 1.: SED models ---|
%==========================|
%
\begin{figure}[th!]
\begin{center}
\resizebox{\hsize}{!}{\includegraphics[angle=-90]{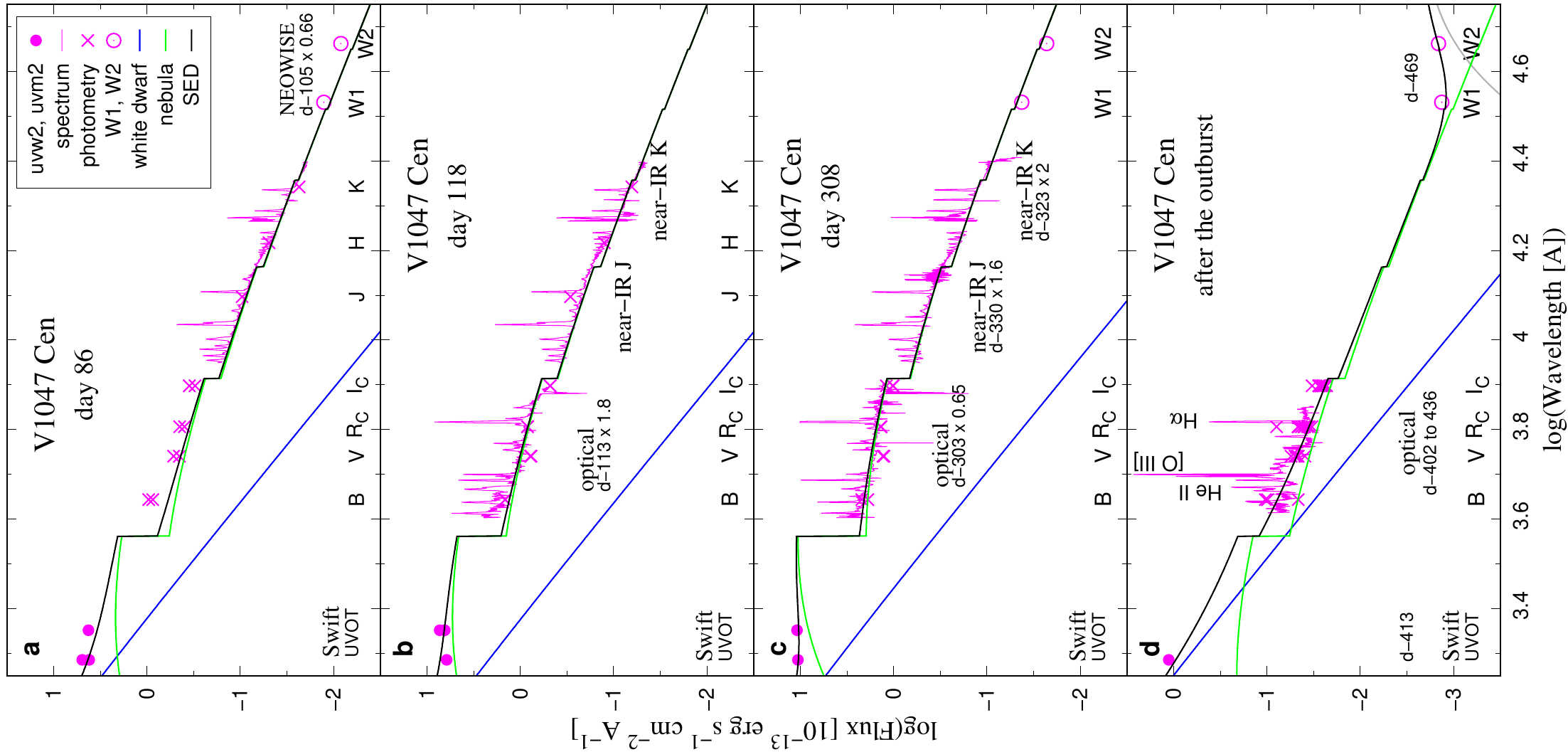}}
%\resizebox{8.5cm}{!}{\includegraphics[angle=-90]{v1047seds4.pdf}}
\end{center}
\caption{
The observed (in magenta) and modeled (black line) SEDs of 
V1047~Cen at selected days (Sect.~\ref{ss:dataforSED}) during 
its 2019 outburst (panels {\bf a}, {\bf b}, and {\bf c}) and 
after it ({\bf d}). The meaning of the lines and symbols is 
shown in the keys. %of panel {\bf a}. 
After the outburst, the NIR $W1$, $W2$ fluxes are affected by 
a third light, interpreted by \cite{2022ApJ...939....6A} as 
a $\sim$400\,K warm dust emission. 
Modeling is described in Sect.~\ref{ss:param}. 
}
\label{fig:seds}
\end{figure}

\subsection{SED modeling and corresponding parameters}
\label{ss:param}
In modeling the NUV to NIR continuum we assumed that the measured 
spectrum, $F(\lambda)$, consists of the stellar component of 
radiation from the hot WD pseudophotosphere, and the nebular 
continuum emitted by the hydrogen plasma. 
The presence of these components of radiation in the spectrum 
is supported by the presence of emission lines of ions with 
a high ionization potential 
\citep[e.g., \civ, \heii, \hi, see][]{2022ApJ...939....6A}. 
Because of a high $T_{\rm WD}$, we observe only the 
long-wavelength tail of the WD's radiation in the NUV/optical, 
which can be approximated by a blackbody radiation. Thus, the 
reddening-free continuum of V1047~Cen measured at the Earth, 
can be written in the form \citep[see][]{2017A&A...604A..48S}
%
%-------- Eq. 1: SED in NUV-NIR --------- 
%
\begin{equation}
 F(\lambda) =   
   \theta_{\rm WD}^2 \pi B_{\lambda}(T_{\rm WD})\,+\,
    k_{\rm N} \varepsilon_{\lambda}({\rm H},T_{\rm e}), 
\label{eq:sed}
\end{equation}
where the angular radius of the WD pseudophotosphere 
$\theta_{\rm WD}$ (= $R_{\rm WD}^{\rm eff}/d$) scales the blackbody 
flux to the spectrum, and it is given by its effective radius 
$R_{\rm WD}^{\rm eff}$ and the distance $d$. 
The observed emission measure, $k_{\rm N}$ (cm$^{-5}$), scales 
the volume emission coefficient 
$\varepsilon_{\lambda}({\rm H},T_{\rm e})$ 
(${\rm erg\,cm^3\,s^{-1}\,\AA^{-1}}$) of the nebular continuum to 
observations: $k_{\rm N}= EM/4\pi d^2$ 
\citep[][]{2005A&A...440..995S} and 
$EM=\int_{V}n_{\rm e}n_{\rm p}{\rm d}V$ is the emission measure of 
the hydrogen nebula given by the electron and proton concentrations, 
$n_{\rm e}$ and $n_{\rm p}$, in the volume $V$. 
The parameter $T_{\rm e}$ is the electron temperature. 

To model the measured SED by Eq.~(\ref{eq:sed}), we first estimated 
the parameters $k_{\rm N}$ and $T_{\rm e}$ from the grid of models 
(\ref{eq:sed}), which determine the amount and the slope of the 
nebular continuum, respectively. 
Second, we added a blackbody radiation for independently determined 
$T_{\rm WD}$ (see below) to fit the NUV fluxes as well. In this way 
we estimated the fitting parameters $\theta_{\rm WD}$, $T_{\rm e}$, 
and $k_{\rm N}$. Consequently, 
$R_{\rm WD}^{\rm eff} = d \theta_{\rm WD}$, 
$EM = 4\pi d^2 k_{\rm N}$, and 
$L_{\rm WD} = 4\pi d^2 \theta_{\rm WD}^2 T_{\rm WD}^4$. 
The SED models are depicted in Fig.~\ref{fig:seds}, resulting 
parameters are listed in Table~\ref{tab:SEDpar} and their temporal 
evolution is shown in Fig.~\ref{fig:lrtw}. 

We estimated the temperature $T_{\rm WD}$ from the 
\heii\,$\lambda$4686/\hb\ flux ratio 
\citep[e.g.,][]{1997pdpn.book.....G}. Here, we used the approach 
of \cite{2020A&A...636A..77S}, who calculated this ratio for 
$T_{\rm e}$ = 30,000 and 20,000\,K, which suits our models 
(Table~\ref{tab:SEDpar}). 
To determine the flux ratio of these lines, we scaled the spectra 
to the emission coefficient 
$\varepsilon_{\lambda}({\rm H},T_{\rm e})$ because the nebular 
continuum dominates the optical/NIR (see Fig.~\ref{fig:seds}). 
The method is valid for $T_{\rm WD}$ between 70,000 and 200,000\,K 
\citep[][]{1981ASIC...69..517I}. 
Table~\ref{tab:EMpar} lists values of $T_{\rm WD}$ for spectra of 
\cite{2022ApJ...939....6A} that contain satisfactorily exposed 
\heii\,$\lambda$4686 and \hb\ emission lines. More details and 
error estimation are given in Appendix~\ref{app:C}. 

Due to the dominance of the nebular continuum in the optical/NIR, 
Eq.~(\ref{eq:sed}) reduces to, $F(\lambda) = 
k_{\rm N} \varepsilon_{\lambda}({\rm H},T_{\rm e})$, which 
allows us determining the parameter $k_{\rm N}$ for any known 
flux-point in this domain. For example, the flux $F(V)$ of 
the emission-line-free $V$ magnitude determines the emission 
measure as 
%
%-------- Eq. 2: EM from F(V) within optical/NIR --------- 
%
\begin{equation}
  EM = 4\pi d^2 \frac{F(V)}{\varepsilon_{V}({\rm H},T_{\rm e})},
\label{eq:em}
\end{equation}
where we used $\varepsilon_{V}({\rm H},30,000\,{\rm K})$ = 3.6 
and $\varepsilon_{V}({\rm H},20,000\,{\rm K})$ = 4.2$\times 10^{-29}$\,
${\rm erg\,cm^3\,s^{-1}\,\AA^{-1}}$ before and after day\,264, 
respectively (see Fig.~\ref{fig:colors}\,a). 
% + 
Having $EM$ and $T_{\rm WD}$ we can determine $L_{\rm WD}$ and 
thus $R_{\rm WD}^{\rm eff}$ under the assumption that all the 
ionizing photons produced by the hot WD pseudophotosphere are 
converted to nebular emission (see Appendix~\ref{app:B} in detail). 
These parameters are introduced in Table~\ref{tab:EMpar} and 
shown in Fig.~\ref{fig:lrtw} as open circles. 
However, after the outburst (day 407), this approach is not 
applicable because the flux of ionizing photons exceeds the rate 
of recombinations required by the measured $EM$ by a factor 
of $>$20.
The uncertainties of all parameters are described in 
Appendix~\ref{app:C}. 
\subsection{Mass-loss rate from the WD}
\label{ss:mdot}
Modeling the broad \ha\ wings, which develop during Z~And-type 
outbursts of SySts, \cite{2006A&A...457.1003S} showed that 
they are emitted by a fast ionized wind flowing from the hot WD 
at rates, ${\dot{M}}_{\rm WD}$, of a few times 
($10^{-7} - 10^{-6}$)\myr. 
Furthermore, the finding that the $EM$ of broad \ha\ wings is 
comparable to that of symbiotic nebulae implies that the 
nebular emission in active SySts can be attributed to the 
emission of the ionized wind from their hot components 
(see Sect.~4.2. there). 
Since the emission lines as well as the nebular continuum arise 
in the same ionized volume, the value of ${\dot{M}}_{\rm WD}$ 
can be determined from the $EM$ and the given wind model. Here, 
we used the relationship between $EM$ and ${\dot{M}}_{\rm WD}$ 
derived by \cite{2017A&A...604A..48S} for a spherically symmetric, 
$\beta$-law wind of \cite{1999isw..book.....L}, 
\begin{equation}
\textsl{EM} = \xi\,
              \Big(\frac{{\dot{M}}_{\rm WD}}{v_{\infty}}\Big)^{2}
              \frac{1}{b R_0 (1-2\beta)}
              \Big[1-\Big(1-\frac{b R_0}{R_{\rm in}}
              \Big)^{1-2\beta}\Big] .
\label{eq:emwind}
\end{equation}
The wind starts at the radial distance $r = R_0$ from the WD 
centre with the initial velocity $a$ and becomes optically thin 
at $r = R_{\rm in} \equiv R_{\rm WD}^{\rm eff}$. 
The wind velocity profile is characterized by the acceleration 
factor $\beta$ and the terminal velocity $v_{\infty}$. 
The parameter $\xi = 1.45\times 10^{46}$\,g$^{-2}$ and 
$b = 1-(a/v_{\infty})^{1/\beta}$. 
We determined ${\dot{M}}_{\rm WD}$ for $EM$ in 
Tables~\ref{tab:SEDpar} and \ref{tab:EMpar}, 
$v_{\infty} = 1700$\kms\ \citep[][]{2022ApJ...939....6A}, 
$\beta = 1.7$ and $a$ = 50\kms\ \citep[][]{2017A&A...604A..48S}. 
For the value of $R_{\rm 0}$, we considered two limiting cases: 
$R_{\rm 0} = R_{\rm WD}^{\rm eff}$ and $R_{\rm 0} = R_{\rm WD}$ 
($\equiv$0.01\ro), which correspond to the lower and upper limits 
of ${\dot{M}}_{\rm WD}$, respectively. 
The averages of these limiting ${\dot{M}}_{\rm WD}$ are given 
in Tables~\ref{tab:SEDpar} and \ref{tab:EMpar} and shown in 
Fig.~\ref{fig:lrtw}\,e. Their uncertainties are described in 
Appendix~\ref{app:C}. 
%
%
%===================================|
%-- Fig. 2: LC + LRTW parameters ---|
%===================================|
%
\begin{figure}[th!]
\begin{center}
\resizebox{\hsize}{!}{\includegraphics[angle=-90]{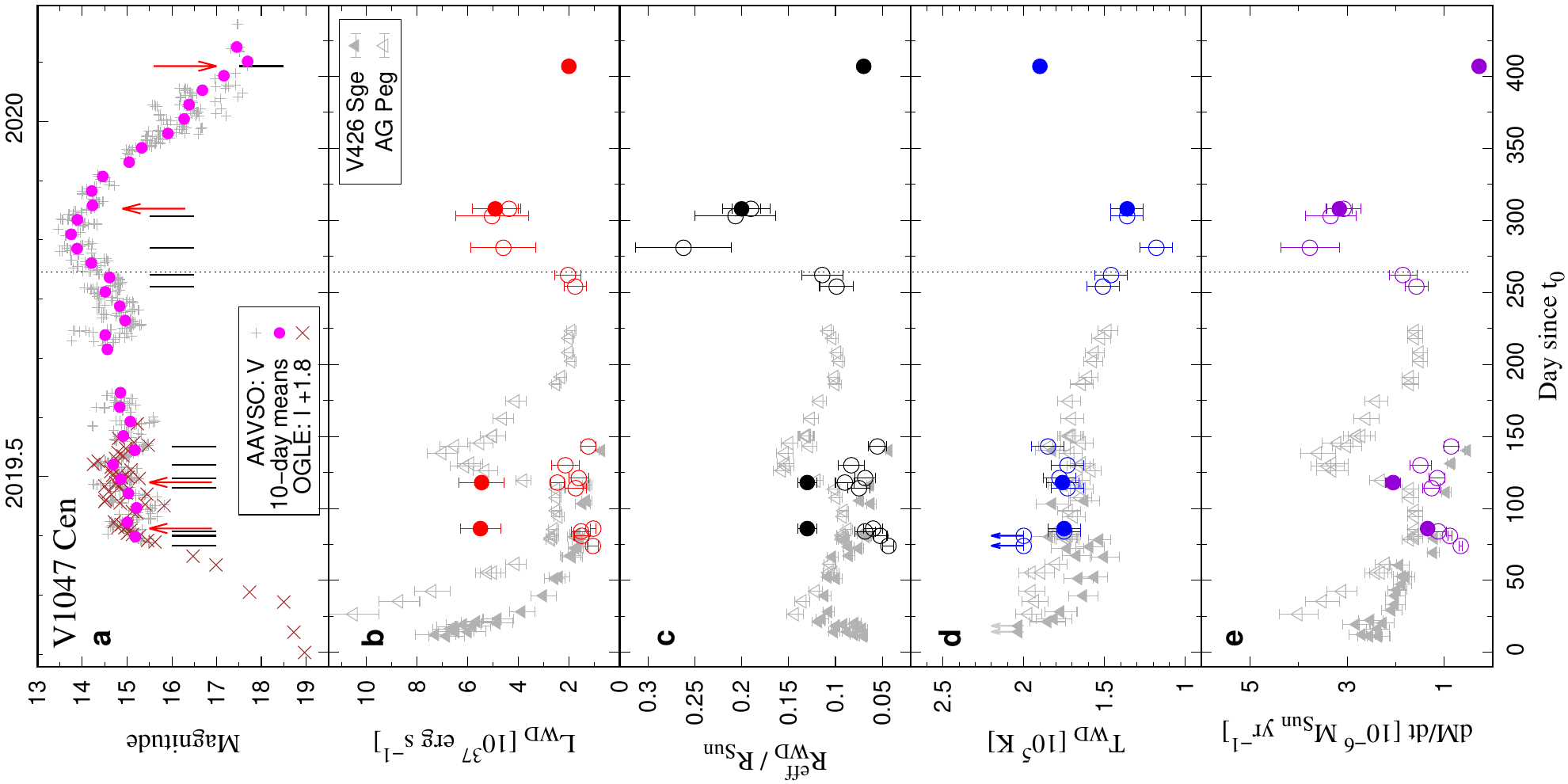}}
%\resizebox{8.5cm}{!}{\includegraphics[angle=-90]{v1047_lrtw.pdf}}
\end{center}
\caption{Evolution of the $V/I$ light curve (panel {\bf a}), and 
parameters $L_{\rm WD}$ ({\bf b}), $R_{\rm WD}^{\rm eff}$ ({\bf c}), 
$T_{\rm WD}$ ({\bf d}), and $\dot M_{\rm WD}$ ({\bf e}) throughout 
the 2019 outburst of V1047~Cen. The full circles in panels 
{\bf b} to {\bf e} are from Table~\ref{tab:SEDpar}, while the open 
circles are from Table~\ref{tab:EMpar}. 
Vertical bars in panel {\bf a} denote the acquisition dates of the 
spectra used to determine $T_{\rm WD}$ (Table~\ref{tab:EMpar}), while 
the red arrows indicate the dates of the SED models (Table~\ref{tab:SEDpar}). 
The vertical dotted line (day 264) indicates the time just before 
the main brightening. 
The gray triangles are for comparison with the values determined 
during Z~And-type outbursts of SySts AG~Peg and V426~Sge 
\citep[data from][]{2017A&A...604A..48S,2020A&A...636A..77S}. 
}
\label{fig:lrtw}
\end{figure}
\subsection{Color-diagram diagnostics}
\label{ss:color}
Here, using the multicolor photometric observations, we independently 
support the main result of the SED modeling -- the presence 
of a strong nebular continuum that dominates the optical/NIR during 
the 2019 outburst of V1047~Cen. 

For this purpose, we reconstructed the reddening-free and line-free 
$V-I_{\rm C}$/time, and $B-V$/$V-I_{\rm C}$ diagrams and compared 
them with the corresponding theoretical diagrams calculated for 
hydrogen nebular continuum emitting at different $T_{\rm e}$. 
The result is shown in Fig.~\ref{fig:colors}, the reconstruction 
of which is described in Appendix~\ref{app:D}. 

From the first observations on day 77 to day $\sim$264 (green 
points), the $V-I_{\rm C}$ index shows a plateau phase with values 
scattered around 0.76\,mag that corresponds to 
$T_{\rm e} \sim $30,000\,K. Relatively stable $T_{\rm e}$ also 
implies a relatively stable temperature of the ionization source 
$T_{\rm WD}$ -- in good agreement with SED models from this period 
(days 86 and 118, Table~\ref{tab:SEDpar}). 
During the following period, from day $\sim$264 to day $\sim$331 
(red points), when the outburst reached its optical maximum, 
the $V-I_{\rm C}$ index has increased to $\sim$1.05\,mag, which 
corresponds to $T_{\rm e}\sim$15,000\,K and suggests a decrease 
in $T_{\rm WD}$ -- also consistent with the SED model performed 
just after the maximum brightness (day 308, Table~\ref{tab:SEDpar}). 
During the decline of the outburst, from day $\sim$343 to day 
$\sim$402 (violet points), the $V-I_{\rm C}$ and $B-V$ indices 
were significantly affected by the appearance of a strong 
nebular [\oiii]\,$\lambda$5007 line in the spectrum 
\citep[][]{2022ApJ...939....6A}, which is located at the steep 
short-wavelength edge of the $V$ filter, and thus can cause 
a large scatter in the photometric measurements obtained by different 
instruments \citep[see Fig.~8 of][]{1993A&A...277..103C}. 
The $B-V$/$V-I_{\rm C}$ diagram (panel {\bf b}) confirms the 
properties of the nebular continuum suggested by the behavior 
of the $V-I_{\rm C}$ index. 

A comparison of the observed and theoretical color indices confirms 
the dominant presence of the nebular continuum in the spectrum and 
changes in its $T_{\rm e}$ as revealed by SED modeling. 
%
%
%==============================|
%-- Fig. 3.: color diagrams ---|
%==============================|
%
\begin{figure}[th!]
\begin{center}
\resizebox{\hsize}{!}{\includegraphics[angle=-90]{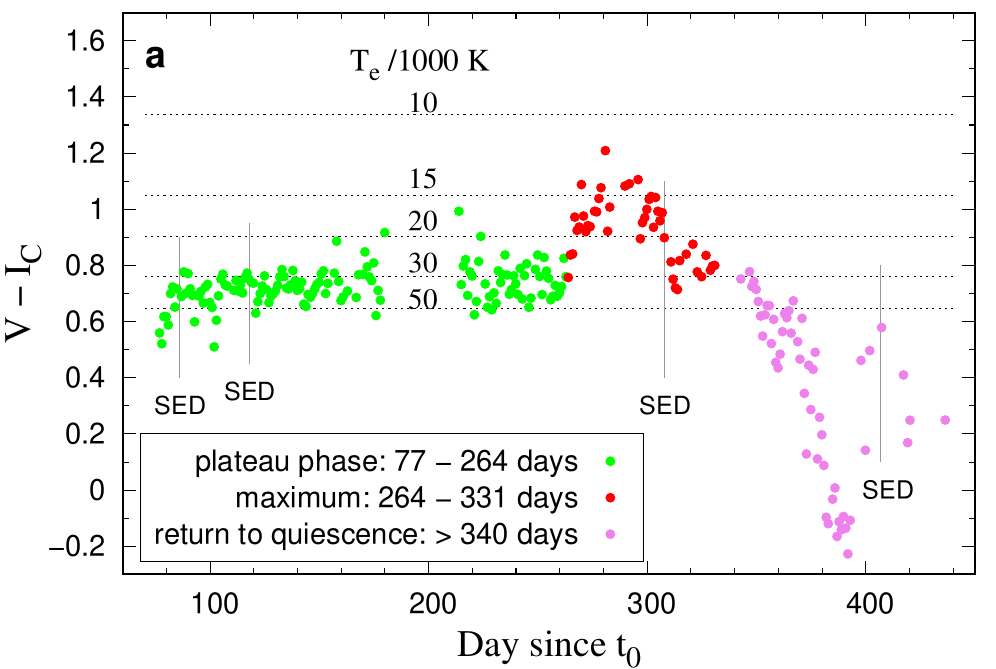}}
\resizebox{\hsize}{!}{\includegraphics[angle=-90]{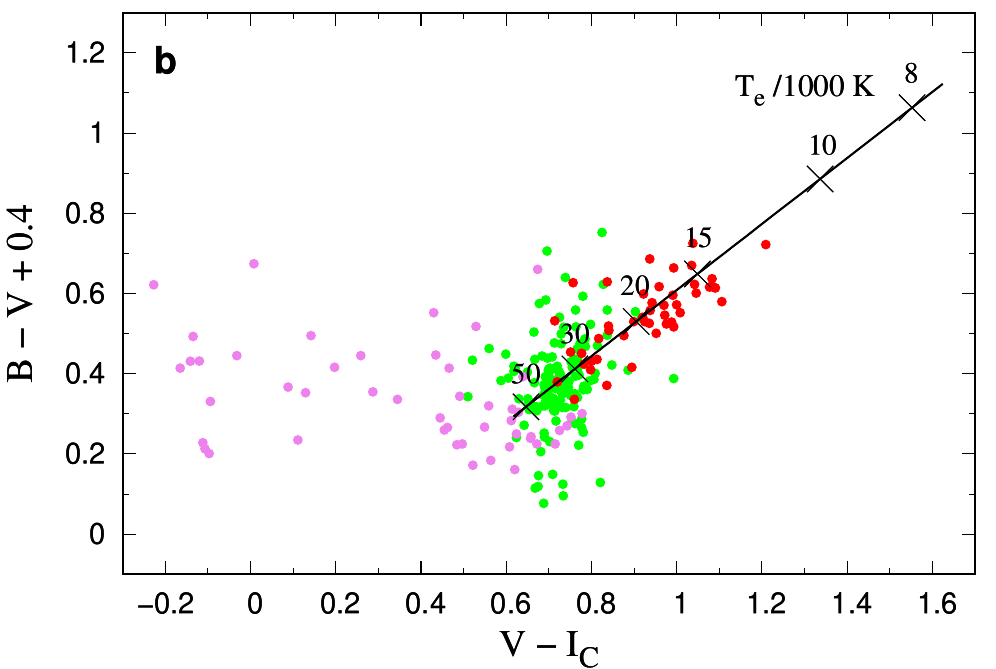}}
\end{center}
\caption{
Color-diagram diagnostics of the 2019 outburst of V1047~Cen. 
The panel {\bf a} shows the evolution of the $V-I_{\rm C}$ color 
index along the outburst, while the panel {\bf b} shows the 
$B-V$/$V-I_{\rm C}$ color diagram. 
Dotted lines in panel {\bf a} and the solid line in panel {\bf b} 
correspond to the theoretical values of the hydrogen nebular continuum 
calculated for different electron temperatures $T_{\rm e}$, marked 
by labels. 
Different colors distinguish specific parts of the light curve 
as indicated in the keys.
The good agreement between the measured and theoretical values 
confirms the dominant contribution of the nebular continuum in 
the optical spectrum, as found by SED modeling 
(see Fig.~\ref{fig:seds}, Sect.~\ref{ss:color}). 
}
\label{fig:colors}
\end{figure}

\subsection{The WD luminosity outside the 2019 outburst}
\label{ss:lumout}
Based on the brightness excess between the 2005 and 2019 outbursts 
(the average OGLE $I$ = 17.1 and $V$ = 17.5\,mag between 2013 and 
2019) compared to the pre-nova outburst ($V\gtrsim 20.5-21)$ and 
the blue reddening-free $V-I$ index, %of -1.19\,mag, 
\cite{2022ApJ...939....6A} concluded that the luminosity of 
the WD before the 2019 outburst was powered by a stable nuclear 
burning on its surface. 
After the 2019 outburst, the authors supported this scenario 
by the presence of high ionization lines in the late spectrum 
of V1047~Cen (days 407, 615, 643, and 774), 
which is a signature of the nuclear burning in a shell on the WD 
surface \citep[e.g.][]{2016acps.confE..21S} and the presence of 
collimated outflow during this period that is also sometimes 
observed in the spectra of SySts at/after their major Z~And-type 
outbursts 
\citep[e.g., Z~And, BF~Cyg, see][]{2009ApJ...690.1222S,
                                   2013A&A...551L..10S}. 

Here, we support this suggestion with a model SED from the 2019 
post-outburst period (Fig.~\ref{fig:seds}\,d), when the brightness 
of V1047~Cen dropped to its pre-outburst value, $V\sim 17.5$\,mag. 
The model shows that the decrease in brightness by $\sim$4\,mag 
in the optical during 100 days after the maximum was mainly due to 
the decrease in $EM$ by a factor of $\sim$40, which corresponds 
to a decrease in the mass-loss rate by a factor of $\gtrsim$10, 
${\dot{M}}_{\rm WD}\sim 2.8\times 10^{-7}$\myr. 
As a result, the optically thick/thin wind interface has shrunk 
to $R_{\rm WD}^{\rm eff}\sim 0.07$\ro, while the increased 
\heii\,$\lambda$4686 emission ($\approx F$(\hb)) indicated a high 
$T_{\rm WD}$ of $\sim$190\,kK, which corresponds to a still high 
$L_{\rm WD}$ of $\sim 2\times 10^{37}$\es\ 
(see Table~\ref{tab:SEDpar}). 
This value is well explained by the nuclear burning of hydrogen-rich 
material on the surface of the WD at a rate of 
$\sim 7.2\times 10^{-8}$\myr\ \footnote{$\dot{M}_{\rm acc} = 
L_{\rm WD}/\eta/X$, where 
$\eta = 6.3\times 10^{18}$\,erg\,g$^{-1}$ is the hydrogen-burning 
energy production from 1 gram, and $X\equiv 0.7$ is the hydrogen 
mass fraction in the accreted matter.} driven by accretion from 
the donor and remnant mass from the 2019 outburst. 

\section{Discussion}
\label{s:discuss}
\subsection{Classification of the 2019 V1047~Cen outburst}
\label{ss:class}
The values of the WD parameters and the wind-mass-loss rate 
during the 2019 V1047~Cen outburst (Table~\ref{tab:SEDpar}, 
Fig.~\ref{fig:lrtw}), and its pre- and post-outburst luminosities, 
which are consistent with stable hydrogen burning on the surface 
of the WD (see Sect.~\ref{ss:lumout}), allow the 2019 outburst 
to be classified as of the Z~And-type, commonly observed in SySts 
(see Sect.~\ref{ss:awdbs}). 

The key to this interpretation is the emergence of strong nebular 
emission in the spectrum during the outburst 
(see Figs.~\ref{fig:seds} and \ref{fig:colors}). Its high value, 
characterized by $EM\gtrsim 10^{60}$\cmt, requires a strong 
ionization source in the system producing $\gtrsim 10^{47}$ 
hydrogen ionizing photons per second, which at the temperature 
of $\sim 1.6\times 10^5$\,K corresponds to the luminosity 
of $\gtrsim 10^{37}$\es\ (see Table~\ref{tab:EMpar}). 
Such a strong ionization source is represented by a nuclear-burning 
WD in V1047~Cen. 

According to the SED models, the WD average luminosity of 
$5.3\times10^{37}$\es\ (see the 6th column in Table~\ref{tab:SEDpar}) 
requires $\dot{M}_{\rm acc} \sim 2\times 10^{-7}$\myr\ throughout 
the 2019 outburst\footnote{i.e., during the main part of the outburst 
covered by multiwavelength observations, from day $\sim$74 to 
day $\sim$340.}, which is above the upper limit of stable burning 
for a WD mass $\lesssim 0.7$\mo\ 
\citep[see e.g., Fig.~1 of][]{2013ApJ...777..136W}. 
Since the nebular emission represents a fraction of the WD radiation 
converted by wind particles, its marked increase during the outburst 
reflects a corresponding increase in the ${\dot{M}}_{\rm WD}$ 
(Sect.~\ref{ss:mdot}). In our case, $EM\gtrsim 10^{60}$\cmt\ 
corresponds to ${\dot{M}}_{\rm WD}\gtrsim 10^{-6}$\myr\ 
throughout the 2019 outburst -- the value proposed by theory when 
the accretion rate on a WD increases above the level sustaining 
stable burning (see Sect.~\ref{ss:awdbs}). 

Therefore, according to the values of the $L_{\rm WD}$ and 
${\dot{M}}_{\rm WD}$ parameters, the 2019 V1047~Cen outburst 
can be classified as a Z~And-type. %(see Sect.~\ref{ss:awdbs}). 

\subsection{The origin of the brightness variability}
\label{ss:dm}
Due to the dominance of the nebular continuum in the optical/NIR 
(Fig.~\ref{fig:seds}), an increase in brightness corresponds to 
a higher $EM$ (Eq.~(\ref{eq:em})), and thus a higher 
${\dot{M}}_{\rm WD}$ (Eq.~(\ref{eq:emwind})), and vice versa. 
Accordingly, variations in the optical light curve on a timescale 
of several days to tens of days, with an amplitude of 
$\lesssim$1\,mag, observed along the 2019 outburst 
\citep[][ Fig.~\ref{fig:lrtw}\,a here]{2022ApJ...939....6A} 
as well as the main brightening after day $\sim$264, can be 
explained by the variable wind from the burning WD. 

The connection between the star's brightness and 
${\dot{M}}_{\rm WD}$ is also supported by the broad emission 
wings of the \hi\ and \hei\ lines ($\sim$2,000\kms) with variable 
P~Cyg type absorption components at the beginning of the first 
flat-peak phase of the outburst 
\citep[see Fig.~3 of][]{2019ApJ...886L..14G}, 
and during the maximum (day $\gtrsim$264), when the enhanced 
${\dot{M}}_{\rm WD}$ (see Fig.~\ref{fig:lrtw}\,e) 
was evidenced by the reappearance of strong P~Cygni profiles in 
the line spectrum \citep[see][]{2022ApJ...939....6A}. 

Therefore, it is natural to assume that the observed variability is 
caused by variable mass transfer from the donor during the outburst, 
which means a variable accretion that is still above the stable 
burning limit (${\dot{M}}_{\rm WD}\gtrsim 10^{-6}$\myr) and hence 
a variable mass-loss rate causing the observed brightness 
variability. 
%
%  additional figure for ApJ-R1
%==================================|
%-- Fig. 4.: L_H & N_rec / time ---|
%==================================|
%
\begin{figure}[t!]
\begin{center}
\resizebox{\hsize}{!}{\includegraphics[angle=-90]{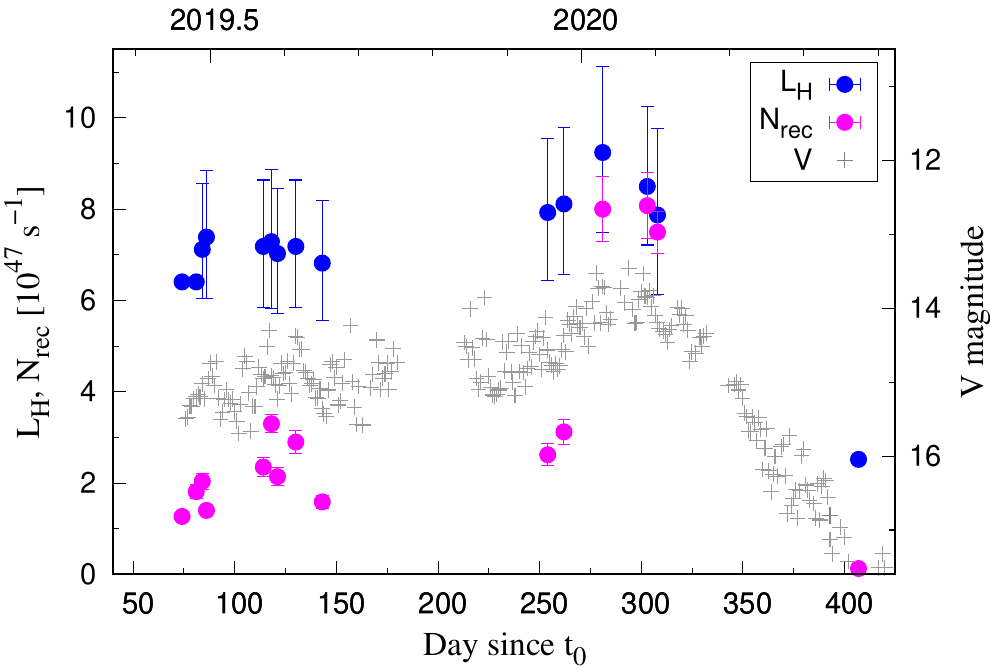}}
\end{center}
\caption{
Evolution of the flux of hydrogen-ionizing photons, $L_{\rm H}$, 
generated by the hot WD pseudophotosphere, and the rate of 
recombinations, $N_{\rm rec}$, in the ionized wind along the 
2019 V1047~Cen outburst (Appendix~\ref{app:B}). The 1-day means 
of $V$ magnitude are compared. The variations in $N_{\rm rec}$ 
reflect those in $EM$, ${\dot{M}}_{\rm WD}$, and hence 
the observed brightness variations (Sect.~\ref{ss:lwd}). 
          }
\label{fig:lph}
\end{figure}

\subsection{The WD luminosity during the 2019 outburst}
\label{ss:lwd}
The determination of $L_{\rm WD}$ from SED models (day 86, 118 
and 308) is given by the temperature, $T_{\rm WD}$, and scaling, 
$\theta_{\rm WD}$, of the WD radiation to the observed NUV to 
NIR SED (Sect.~\ref{ss:param}). Despite the significantly 
different $T_{\rm WD}$ (Table~\ref{tab:EMpar}) and 
$\theta_{\rm WD}$ between the first part of the outburst (days 
86 and 118) and around the maximum (day 308), the corresponding 
$L_{\rm WD}$ is stable (within the uncertainties), with an 
average value of $5.3\times 10^{37}$\es, throughout the outburst 
(Table~\ref{tab:SEDpar}, Fig.~\ref{fig:lrtw}\,b). 
%

% L_WD from EM: new paragraph
The estimate of $L_{\rm WD}$ from the nebular continuum assumes 
an equilibrium between the flux of hydrogen-ionizing photons, 
$L_{\rm H}$, generated by the hot WD pseudophotosphere and 
the rate of recombinations, $N_{\rm rec}$, in the wind 
(see Appendix~\ref{app:B}). 
Figure~\ref{fig:lph} shows that values of $L_{\rm H}$ are 
comparable (within the uncertainties) during the outburst, 
while values of $N_{\rm rec}$ correlate with the star's brightness. 
The former is a result of a constant $L_{\rm WD}$ at high 
$T_{\rm WD}$ (see Eq.~(\ref{eq:lh}), while the latter is given 
by variations in the $EM$ (according to Eqs.~(\ref{eq:nrec}) 
and (\ref{eq:em}), $V = \log(N_{\rm rec})$ + const.). 

During the first part of the 2019 outburst (days $<264$), 
$L_{\rm H} > N_{\rm rec}$. This means that a fraction of 
ionizing photons escapes the nebula without being converted to 
nebular radiation. In other words, the wind particles are unable 
to convert all ionizing photons into nebular radiation. 
As a result, the corresponding $EM$ provides values of 
$L_{\rm WD}$ according to Eq.~(\ref{eq:lwdem}) (open circles 
in Fig.~\ref{fig:lrtw}\,b), which are smaller than the true 
values (full circles in the figure). 

During the maximum, $L_{\rm H}\sim N_{\rm rec}$, which means that 
all ionizing photons are absorbed by the wind particles and converted 
into nebular radiation -- we measure a maximum $EM$ in the 
spectrum for a given $L_{\rm H}$. As a result, $L_{\rm WD}$ 
obtained from both, $EM$ (Eq.~(\ref{eq:lwdem}) and SED models 
are comparable (see Fig.~\ref{fig:lrtw}\,b). 

\subsection{Consistency with the optically thick wind theory}
\label{ss:consist}
%
%  For the outburst interpretation, it is important  that ...
The nature of the 2019 outburst of the Nova V1047~Cen also 
dictates the mutual consistency of the WD parameters, 
$L_{\rm WD}$, $R_{\rm WD}^{\rm eff}$, 
$T_{\rm WD}$, ${\dot{M}}_{\rm WD}$, and $T_{\rm e}$ of the ionized 
wind: A higher ${\dot{M}}_{\rm WD}$ creates a larger optically 
thick/thin wind interface above the WD surface (i.e., a larger 
$R_{\rm WD}^{\rm eff}$) resulting in a lower $T_{\rm WD}$ that is 
responsible for a lower $T_{\rm e}$, and vice versa. 
The parameters we obtained during the first flat-peak phase of 
the outburst and during its maximum well express this relationship 
(see Fig.~\ref{fig:lrtw}, Sect.~\ref{ss:color}). 
This is consistent with the theory of the optically thick wind 
in outbursts on the WD surface \citep[see][]{1994ApJ...437..802K}. 

\subsection{Comparison with Aydi et al. (2022)}
\label{ss:aydi}
Although DN outbursts occasionally occur in CN binaries, 
observations of the 2019 outburst in the CN V1047~Cen were 
controversial in this regard 
\citep[see][]{2019ApJ...886L..14G}. %,2022ApJ...939....6A}. 
Also, \cite{2022ApJ...939....6A} questioned the possibility of 
a DN at the beginning of the 2019 outburst due to the high 
accretion rate and heating of the accretion disk during 
the stable burning phase between the 2005 and 2019 outbursts. 
Based on the multiwavelength observations, 
\cite{2022ApJ...939....6A} came to a conclusion that this event 
may have started with a brightening of the disk, which then 
triggered enhanced nuclear shell burning on the WD surface. 
The authors also pointed out that the 2019 outburst had many 
observational similarities to outbursts of classical SySts. 
However, without deriving correct parameters (compare their SED 
model in Fig.~14 and our one in Fig.~\ref{fig:seds} here), the 
authors were unable to clearly classify the outburst, even 
admitting the possibility of a {\it "record breaking dwarf nova 
outburst or a new phenomenon"}. 

Our classification of the 2019 V1047~Cen outburst as 
a Z~And-type is given by the following results: 
%
%\begin{enumerate}
\begin{itemize}
\item
Before and after the outburst, $L_{\rm WD}$ corresponded to 
values generated by a stable-burning WD 
(see Sect.~\ref{ss:lumout}). 
\item
During the outburst (from day $\sim$74 to $\sim$340), the high 
$L_{\rm WD}\sim 5.3\times 10^{37}$\es\ requires fueling at 
$\dot{M}_{\rm acc}\sim 2\times 10^{-7}$\myr\ that is above 
the stable burning limit for a WD mass $\lesssim 0.7$\mo. 
\item 
At the same time, ${\dot{M}}_{\rm WD}\gtrsim 10^{-6}$\myr\ 
was indicated (see Sect.~\ref{ss:class}). 
\end{itemize}
%\end{enumerate}
%
These properties constitute the basic conditions for classifying 
the outburst as a Z~And-type (see Sect.~\ref{ss:awdbs}). 

In agreement with the above-mentioned papers, there is no 
reason for the appearance of a DN outburst during this main 
part of the 2019 outburst. 
Also, the gradual brightening during the first two months after 
the outburst onset suggests a gradual increase in the accretion 
throughout the disk, and not a one-time brightening due to disk 
instability since the profile of the ascending part of the 2019 
outburst does not match that of a DN outburst. 
Accordingly, it is possible to imagine that a gradual increase 
in the mass transferred through the disk, will cause the disk, 
and consequently, the WD, gradually brightens. 
Since the nuclear hydrogen burning is a factor of $\sim$50 more 
efficient than the release of radiation by the accretion process, 
it is possible to conclude that the initial brightening by 
$\gtrsim$4\,mag is the result of the gradual feeding of a stable 
burning WD to the upper limit, at/above which the unconsumed mass 
begins to flow away. 

\section{Conclusion}
\label{s:concl}
In this work, we have shown that the outburst in the CN V1047~Cen 
(Nova Cen 2005), which appeared in 2019, is of the Z~And-type -- a 
type that has so far been observed exclusively in the light curves 
of SySts (Sect.~\ref{ss:class}). 

The 2019 outburst was powered by the accretion of the hydrogen-rich 
material on the surface of a steady-burning WD at rates of 
$\sim 2\times 10^{-7}$\myr\ that is above the upper limit of 
stable burning for a WD mass $\lesssim 0.7$\mo. 
Following the theory, this led to the formation of a stellar wind 
blowing at rates $\gtrsim 10^{-6}$\myr\ (Sects.~\ref{ss:mdot} and 
\ref{ss:class}). 
The wind was ionized by a $\sim 1.6\times 10^5$\,K hot and 
$\sim 5\times 10^{37}$\es\ luminous, nuclear-burning WD giving 
rise to a large amount of nebular radiation 
($EM\gtrsim 10^{60}$\cmt) that dominated the optical/NIR 
(Figs.~\ref{fig:seds} and \ref{fig:lrtw}, Tables~\ref{tab:SEDpar} 
and \ref{tab:EMpar}). 
These characteristics are typical of Z~And-type outbursts in 
SySts (see Sect.~\ref{ss:awdbs}, and Sect.~\ref{ss:aydi} for 
summary). 

The presence of the Z~And-type outburst in the CN binary 
indicates an extension of the nuclear burning time on the surface 
of the WD after the CN explosion. This may lead to a faster 
mass increase of the accreting WD, and thus to a faster 
evolution terminated by Type Ia supernova explosion. 
However, to better understand this possibility in the evolution 
of CVs, we need to know the nature of the donor star and how 
a nova explosion can affect it. 

\begin{acknowledgments}
The authors thank the anonymous referee for constructive comments. 
This work was supported by a grant of the Slovak Academy of 
Sciences, VEGA No. 2/0003/25 and by the Slovak Research and 
Development Agency under the contract No. APVV-20-0148. 
We also acknowledge the variable-star observations from the 
AAVSO International Database contributed by observers worldwide 
and used in this research. 
\end{acknowledgments}
\clearpage
\appendix
\section{Magnitudes of the true continuum}
\label{app:A}
Photometric magnitudes represent integrated fluxes that include 
both the continuum and the line spectrum. The presence of emission 
lines leads to brighter magnitudes than those of the continuum. 
The influence of an emission line on the measured magnitude, 
$m_{\rm obs}$, is given by its equivalent width, the width of 
the photometric filter, and its transmission at the wavelength of 
the line. 
According to the method of \cite{2007NewA...12..597S}, we determined 
corrections $\Delta m = m_{\rm obs} - m_{\rm cont}$ for emission 
lines, where $m_{\rm cont}$ is the magnitude of the true continuum. 
Using the spectra of \cite{2022ApJ...939....6A}, we obtained 
corrections $\Delta B$, $\Delta V$, and $\Delta R_{\rm C}$ 
(Table~\ref{tab:dml}). 
The values of $\Delta B$ represent their lower limit since 
the spectra cover only part of the $B$ filter. Similarly, 
corrections $\Delta I_{\rm C}$ could be estimated only for 
days 74, 81 ($\Delta I_{\rm C} \gtrsim -0.05$\,mag), and 297, 303 
($\Delta I_{\rm C}\approx -0.1$\,mag, see Appendix~\ref{app:D}). 
In modeling the SED after the outburst, we assumed 
$\Delta I_{\rm C} = -0.1$\,mag. 
Magnitudes of the true continuum were used for the flux calibration 
of the optical spectra (Sect.~\ref{ss:dataforSED}) and for the 
color-diagram diagnostics (Sect.~\ref{ss:color}). 

%
%=====================================|
%---- Table A.1: Corrections Dm_l ----|
%=====================================|
%
\begin{table}[h]
\caption{Corrections of the $B$, $V$, and $R_{\rm C}$ magnitudes 
         for emission lines measured on the spectra of 
         \cite{2022ApJ...939....6A}. %, see Appendix~\ref{app:A}. 
}
\begin{center}
\begin{tabular}{lccccc}
\hline
\hline
\noalign{\smallskip}
Age$^{\rm a}$ 
        &Specrum range& $\Delta B$   & $\Delta V$   & $\Delta R_{\rm C}$ \\
(days)  &  (nm)       &   (mag)      &   (mag)      &   (mag)   \\
\hline
\noalign{\smallskip}
~~74    &410--880     &$<$ -0.06$^b$ &   -0.07      &   -0.12   \\
~~81    &400--880     &$<$ -0.26     &   -0.08      &   -0.14   \\
~~84    &450--520     &$<$ -0.21$^b$ &    --        &     --    \\
114     &390--560     &$<$ -0.19$^b$ &$<$ -0.04     &     --    \\
121     &380--780     &    -0.26     &   -0.08      &   -0.23   \\
130$^b$ &450--890     &      --      &   -0.09      &   -0.22   \\
143     &380--780     &    -0.33     &   -0.08      &   -0.29   \\
231     &485--780     &      --      &   -0.09      &   -0.22   \\
254     &380--780     &$<$ -0.21$^b$ &   -0.12      &   -0.32   \\
262     &380--780     &$<$ -0.16$^b$ &   -0.13      &   -0.28   \\
281     &450--820     &$<$ -0.10$^b$ &   -0.12      &   -0.31   \\
286     &450--820     & --           &   -0.08      &   -0.21   \\
297     &420--890     &$<$ -0.11$^b$ &   -0.13      &   -0.34   \\
303     &400--880     &$<$ -0.16$^b$ &   -0.10      &   -0.27   \\
407     &410--710     &$<$ -0.43     &   -0.40      &   -0.25   \\
\hline
\end{tabular}
\end{center}
{\bf Notes.} \\
$^a$ As in Table~\ref{tab:SEDpar}. \\
$^b$ Underexposed spectrum. 
\label{tab:dml}
\end{table}
%

%
%=============================================================|
%-- Table B.1: Age, Twd, V, EM, Nrec, Lwd, LH,  Rwd, dotMwd --|
%=============================================================|
%
\begin{table}
%\tiny
%\scriptsize
%\footnotesize
%\small
\caption{Physical Parameters of the 2019 V1047~Cen Outburst 
Determined with the Aid of Emission Measure$^{\rm a}$. 
}
%\vspace{-6mm}
\begin{center}
\begin{tabular}{ccccccccc}
\hline
\hline
\noalign{\smallskip}
Age$^{\rm b}$                      &
$T_{\rm WD}$                       &
$V^{\rm d}$                        &
$EM$                               &
$N_{\rm rec}$                      &
$L_{\rm WD}$                       &
$L_{\rm H}$                        &
$R_{\rm WD}^{\rm eff}$             &
${\dot{M}}_{\rm WD}$               \\
%--------------------------
(days)                             &
(kK)                               &
(mag)                              &
(10$^{60}$\,cm$^{-3}$)             &
($10^{47}{\rm s^{-1}}$)            &
($10^{37}{\rm erg\,s^{-1}}$)       & 
($10^{47}{\rm s^{-1}}$)            &
($R_{\odot}$)                      &
($10^{-6}M_{\odot}\,{\rm yr}^{-1}$)\\
\noalign{\smallskip}
\hline
\noalign{\smallskip}
%Age |    Twd   |  V   |           EM            |     Nrec     |     Lwd      |      Lph              |    Rwd        | dM/dt        \\
~~74 &$>$200    &15.58 & 1.27$\pm$0.11           &1.27$\pm$0.11 & $>$1.05      &$\sim$6.41             & $\sim$0.044   & $\sim$0.7    \\
~~81 &$>$200    &15.18 & 1.81$\pm$0.16           &1.81$\pm$0.16 & $>$1.50      &$\sim$6.41             & $\sim$0.052   & $\sim$0.9    \\ 
~~84 &175$\pm$10&15.06 & 2.04$\pm$0.18           &2.04$\pm$0.18 & 1.52$\pm$0.37&$7.12_{-1.07}^{+1.44}$ & 0.07$\pm$0.01 &1.12$\pm$0.15 \\
~~86 &175$\pm$10&15.20 & 1.40$\pm$0.10$^{\rm c}$ &1.40$\pm$0.10 & 1.04$\pm$0.12&$7.39_{-1.34}^{+1.46}$ & 0.06$\pm$0.01 &0.85$\pm$0.12 \\
 114 &173$\pm$10&14.94 & 2.35$\pm$0.21           &2.35$\pm$0.21 & 1.73$\pm$0.42&$7.18_{-1.34}^{+1.46}$ & 0.08$\pm$0.01 &1.26$\pm$0.18 \\
 118 &176$\pm$10&14.92 & 3.30$\pm$0.20$^{\rm c}$ &3.30$\pm$0.20 & 2.46$\pm$0.26&$7.29_{-1.46}^{+1.59}$ & 0.09$\pm$0.01 &1.70$\pm$0.17 \\
 121 &178$\pm$10&15.00 & 2.14$\pm$0.19           &2.14$\pm$0.19 & 1.61$\pm$0.39&$7.03_{-1.31}^{+1.42}$ & 0.07$\pm$0.01 &1.14$\pm$0.17 \\
 130 &173$\pm$10&14.66 & 2.90$\pm$0.26           &2.90$\pm$0.26 & 2.14$\pm$0.54&$7.18_{-1.34}^{+1.46}$ & 0.08$\pm$0.01 &1.50$\pm$0.23 \\
 143 &185$\pm$10&15.32 & 1.59$\pm$0.14           &1.59$\pm$0.14 & 1.24$\pm$0.30&$6.82_{-1.27}^{+1.37}$ & 0.06$\pm$0.01 &0.86$\pm$0.15 \\
 254 &151$\pm$10&14.57 & 2.62$\pm$0.24           &2.62$\pm$0.24 & 1.75$\pm$0.44&$7.93_{-1.50}^{+1.63}$ & 0.10$\pm$0.02 &1.57$\pm$0.24 \\
 262 &146$\pm$10&14.37 & 3.12$\pm$0.28           &3.12$\pm$0.28 & 2.04$\pm$0.51&$8.11_{-1.53}^{+1.67}$ & 0.11$\pm$0.02 &1.85$\pm$0.29 \\
 281 &118$\pm$10&13.82 & 5.23$\pm$0.47           &8.00$\pm$0.72 & 4.59$\pm$1.29&$9.24_{-1.75}^{+1.88}$ & 0.26$\pm$0.06 &3.77$\pm$0.61 \\
 303 &136$\pm$10&13.83 & 5.28$\pm$0.47           &8.08$\pm$0.72 & 5.04$\pm$1.44&$8.50_{-1.28}^{+1.75}$ & 0.21$\pm$0.04 &3.34$\pm$0.53 \\
 308 &136$\pm$10&14.20 & 4.90$\pm$0.30$^{\rm c}$ &7.50$\pm$0.46 & 4.37$\pm$0.47&$7.87_{-1.74}^{+1.89}$ & 0.19$\pm$0.02 &3.07$\pm$0.35 \\
% 407 &$\sim$190 &17.69 & 0.13$\pm$0.02           &0.13$\pm$0.02 &$\sim 0.11^{\rm e}$ &$\sim$2.52 & $\sim 0.02^{\rm e}$ &$\sim 0.1^{\rm e}$  \\
%\noalign{\smallskip}
\hline
\end{tabular}
\end{center}
\normalsize
{\bf Notes.} \\ 
$^{\rm a}$ Designation as in Sections~\ref{ss:param}, \ref{ss:mdot}, 
           and Appendix~\ref{app:B}. 
           $EM$ from $V$ according to Eq.~(\ref{eq:em}) or 
                Table~\ref{tab:SEDpar}$^{\rm c}$; 
           $L_{\rm WD}$ from $T_{\rm WD}$ and $EM$ according to 
                  Eq.~(\ref{eq:lwdem}); 
           $R_{\rm WD}^{\rm eff}$ from the Stefan-Boltzmann law; 
           ${\dot{M}}_{\rm WD}$ from Eq.~(\ref{eq:emwind}); 
           $N_{\rm rec}$ from Eq.~(\ref{eq:nrec}), and 
           $L_{\rm H}$ from Eq.~(\ref{eq:lh}). 
           $T_{\rm WD}$ was determined from optical spectra available 
           at given dates (Sect.~\ref{ss:param}). \\
$^{\rm b}$ As in Table~\ref{tab:SEDpar}, \\
$^{\rm c}$ $EM$ from SED models (Table~\ref{tab:SEDpar}), \\
$^{\rm d}$ Observed $V$ magnitude from Fig.~\ref{fig:lrtw}\,a. 
\label{tab:EMpar}
\end{table}

\section{Relationship between $L_{\rm WD}$ and $EM$}
\label{app:B}
The enhanced mass outflow during outbursts converts more ionizing 
radiation from the WD into nebular emission than during quiescence 
(Sects.~\ref{ss:awdbs} and \ref{ss:mdot}). 
In the limiting case, when all the ionizing photons are converted 
into the nebular emission, we can estimate the WD luminosity, 
$L_{\rm WD}$, from the emission measure, $EM$, for the given WD 
temperature, $T_{\rm WD}$, according to the expression 
\citep[see Eq.~(11) of][]{2017A&A...604A..48S}, 
\begin{equation}
  L_{\rm WD} = \alpha_{\rm B}({\rm H},T_{\rm e})\,\textsl{EM}
               \frac{\sigma T_{\rm WD}^{4}}{f(T_{\rm WD})},   
\label{eq:lwdem}
\end{equation}
where $\alpha_{\rm B}({\rm H},T_{\rm e})$ is the total recombination 
coefficient to all but the ground state of hydrogen 
(i.e., describing a dense nebula, optically thick in 
the Lyman continuum; the so-called Case~$B$). 
The corresponding radius, $R_{\rm WD}^{\rm eff}$, is determined 
by the Stefan-Boltzmann law. 
The function $f(T_{\rm WD})$ determines the flux of ionizing 
photons emitted by 1\,cm$^2$ area of the blackbody source 
(cm$^{-2}$\,s$^{-1}$) and can be expressed as 
\begin{equation}
  f(T_{\rm WD}) = \frac{\pi}{hc} \int^{\rm 912\AA}_{0}\!\! \lambda B_{\lambda}
                  (T_{\rm WD})\,{\rm d}\lambda .
\label{eq:fth}
\end{equation}
The total flux of hydrogen ionizing photons, $L_{\rm H}$ (s$^{-1}$), 
emitted by the WD pseudophotosphere is 
\citep[see Eq.~(11) of][]{2001A&A...366..157S}
\begin{equation}
   L_{\rm H}= \frac{L_{\rm WD}}{\sigma T_{\rm WD}^{4}}\,f(T_{\rm WD}) .
\label{eq:lh}
\end{equation}
Assuming the hydrogen plasma, the coefficient 
$\alpha_{\rm B}({\rm H},T_{\rm e})$ expresses the ability of 
1 electron and 1 proton in 1\,cm$^3$ to recombine in 1 second 
(cm$^3$\,s$^{-1}$). Accordingly, physical meaning of 
the parameters $EM$ (see Sect.~\ref{ss:param}) and 
$\alpha_{\rm B}({\rm H},T_{\rm e})$ determine the rate of 
recombinations within the nebule (i.e., the ionized wind), 
$N_{\rm rec}$ (s$^{-1}$), as 
\begin{equation}
   N_{\rm rec} = \alpha_{\rm B}({\rm H},T_{\rm e})\times EM ,
\label{eq:nrec}
\end{equation}
which is valid for the hydrogen plasma, heated by photoionizations. 

In the case, when the outflowing particles are capable of 
converting only a part of the ionizing photons into the nebular 
emission, i.e., $L_{\rm H} > N_{\rm rec}$ (the so-called 
particle-bounded nebula), the value of $L_{\rm WD}$ according to 
Eq.~(\ref{eq:lwdem}) is below its real value. 
Under this condition, increasing ${\dot{M}}_{\rm WD}$ increases 
$N_{\rm rec}$, i.e., $EM$, and thus the brightness of the star, 
as the nebular continuum dominates the optical/NIR 
(Fig.~\ref{fig:seds}, Eq.~(\ref{eq:em})). 
When $L_{\rm H} = N_{\rm rec}$ (the so-called ionization-bounded 
nebula), all the ionizing photons are absorbed within the nebula 
and converted to the nebular emission producing a maximum $EM$. 
As a result, the value of $L_{\rm WD}$ from Eq.~(\ref{eq:lwdem}) 
corresponds to the true value obtained from SED models (see 
Figs.~\ref{fig:lrtw}\,b and \ref{fig:lph} during 
the maximum)\footnote{
Possible brightness increase due to the stellar radiation from 
the hot WD is very small in the optical/NIR, but can be more 
pronounced in the far-UV}. 

All physical parameters obtained with the aid of $EM$, estimated from 
(dereddened and emission-line-free) $V$ magnitudes (Eq.~(\ref{eq:em})) 
and determined by SED models (Eq.~(\ref{eq:sed})), as well as 
the corresponding auxiliary parameters $N_{\rm rec}$ 
(Eq.~(\ref{eq:nrec})) and $L_{\rm H}$ (Eq.~(\ref{eq:lh})) 
are introduced in Table~\ref{tab:EMpar}. 
In determining $N_{\rm rec}$ we used the coefficient 
$\alpha_{\rm B}({\rm H},30,000\,{\rm K})$ = 1.0 and 
$\alpha_{\rm B}({\rm H},20,000\,{\rm K})$ = 
1.53$\times 10^{-13}$\,cm$^{3}$\,s$^{-1}$ 
for electron concentration 10$^{10}$\cmt\ 
\citep[][]{1987MNRAS.224..801H}
during the first part of the 2019 outburst (day $<$ 264) and 
during its maximum, respectively. 
In determining $L_{\rm H}$ we used $L_{\rm WD}$ obtained by 
SED modeling on days 86, 118, and 308 (Table~\ref{tab:SEDpar}), 
and their average value $(5.3\pm 0.8)\times10^{37}$\es\ for 
other days with available $T_{\rm WD}$. 
%x After the outburst, on day 407, we used 
%x $L_{\rm WD} = 2\times 10^{37}$\es, given by SED model. 
The values of $N_{\rm rec}$ and $L_{\rm H}$ along the outburst 
are plotted in Fig.~\ref{fig:lph} and discussed in 
Sect.~\ref{ss:lwd}. 
Their uncertainties are described in Appendix~\ref{app:C}. 

\section{Estimate of uncertainties}
\label{app:C}
% Twd:
The uncertainty of the temperature $T_{\rm WD}$ is given by the 
accuracy of determining the \heii\,$\lambda$4686/\hb\ flux ratio, 
which depends on the $S/N$ ratio and the blending of the \heii\ 
line with the C\,{\scriptsize III}/N\,{\scriptsize III} emissions 
at 4640--4650\,\AA, eventually with the \hei\,$\lambda$4713 
emission line. We estimated the $S/N$ ratio at 5-10\% and 
eliminated the blending by extracting the \heii\ line using 
Gaussian functions. 
The advantage is that the line flux ratio reduces these errors. 
For higher $T_{\rm WD}$, the method transfers a given error in 
the line ratio to a larger uncertainty in $T_{\rm WD}$, and 
vice versa \citep[see Fig.~5.1. of][]{1997pdpn.book.....G}. 
However, smaller values of the line ratio, which 
correspond to a lower $T_{\rm WD}$, are usually burdened by 
a larger error of the \heii\,$\lambda$4686 line flux. 
For spectra with a complicated continuum profile around 
\heii\,$\lambda$4686 and \hb\ lines (days 74, 81, 130 and 407), 
instead of the flux ratio, we used the ratio of their equivalent 
widths, which is a source of additional errors. 
Taking all these limitations into account, we estimated errors 
in $T_{\rm WD}$ of less than 10\%. Since their exact determination 
is difficult, we adopted an average uncertainty in $T_{\rm WD}$ 
of 10,000\,K for all values. 

% theta:
To estimate the uncertainty of the angular radius, 
$\theta_{\rm WD}$, we scaled the Planck function for 
$T_{\rm WD}\pm$10,000\,K to the mean values of NUV fluxes, 
because they fit well with the overall SED models. 
In this way, we obtained uncertainties of $\sim$4\% in 
$\theta_{\rm WD}$ and thus also in $R_{\rm WD}^{\rm eff}$. 
Corresponding uncertainties in $L_{\rm WD}$ that include 
also those in $T_{\rm WD}$ are of 15--20\%\ 
(see Table~\ref{tab:SEDpar}). 

% EM & Te from SED models:
The uncertainty of the scaling factor, $k_{\rm N}$ and thus 
of the $EM$, is given by the accuracy of the optical/NIR 
continuum, which was estimated at $\sim$7\%. 
The electron temperature $T_{\rm e}$ determines the slope of 
the nebular continuum. For the SED models, we estimated the 
maximum error in $T_{\rm e}$ to be about 15\%. 

% EM from V magnitudes, Eq. 2. 
To estimate uncertainties in the $EM$ that was obtained from the 
$V$ magnitude, we adopted the errors of the daily $V$-magnitude 
averages to 0.1\,mag, which translates into uncertainties of 
$\sim$9\%\ in the $EM$ (see Eq.~(\ref{eq:em})). 
% Lwd from Eq. B.1 and corresponding Rwd
Having uncertainties in $EM$ and $T_{\rm WD}$, we determined 
the uncertainty in $L_{\rm WD}$ as the root mean square error 
using the total differential of the function given by 
Eq.~(\ref{eq:lwdem}). In this way, we obtained errors in 
$L_{\rm WD}$ between 24 and 28\%. 
Similarly, we determined uncertainties for the effective 
radius $R_{\rm WD}^{\rm eff}$ given by the Stefan-Boltzman law 
for the errors in $L_{\rm WD}$ and $T_{\rm WD}$. The errors of 
the parameters obtained indirectly using Eqs.~(\ref{eq:em}) and 
(\ref{eq:lwdem}) are given in Table~\ref{tab:EMpar}. 

% uncertainty in dM/dt: 
To estimate the uncertainties in ${\dot{M}}_{\rm WD}$, we followed 
the theory of the optically thick wind in nova outbursts, according 
to which the matter is accelerated deep inside the pseudophotosphere 
\citep[][]{1994ApJ...437..802K}. 
Since the pseudophotosphere represents the optically thick/thin 
interface of the expanding wind, the wind cannot start there. 
However, the onset of the wind is not specified by theory, nor 
can it be determined from observations. Therefore, we adopted 
the average value from both limiting cases as a reasonable 
estimate of ${\dot{M}}_{\rm WD}$ (see Sect.~\ref{ss:mdot}), 
whose maximum error includes the uncertainty of $EM$ and 
$R_{\rm WD}^{\rm eff}$. This gives uncertainties in 
${\dot{M}}_{\rm WD}$ of around 8\%\ in Table~\ref{tab:SEDpar} 
and 13 to 17\%\ in Table~\ref{tab:EMpar}. 

% uncertainty in L_H and Nrec:
For $L_{\rm H}$ (Eq.~(\ref{eq:lh})), we determined the maximum 
error that correspond to the uncertainties in $L_{\rm WD}$ and 
the different values of the $f(T_{\rm WD})$ function for 
the lower and upper limits of a given $T_{\rm WD}$. The error 
values are in the range of 15 to 20\%. Since the $f(T_{\rm WD})$ 
function decreases unevenly with $T_{\rm WD}$ 
(for $T_{\rm WD} > $73,000\,K), we determined the lower and 
upper error limits in $L_{\rm H}$ separately. 
Finally, the uncertainties in $N_{\rm rec}$ reflect those 
in $EM$ (see Eq.~(\ref{eq:nrec})). 

\section{Reconstruction of color diagrams}
\label{app:D}
Using data from the AAVSO database from day 77 to day 407, we 
determined the $B-V$ and $V-I_{\rm C}$ indices from the average 
values of $B$, $V$, and $I_{\rm C}$ magnitudes observed within 
1 day. In this way, we obtained the color indices measured in 
238 days along the 2019 outburst of V1047~Cen. 
We dereddened the magnitudes with $E_{\rm B-V} = 1.0$\,mag 
using the extinction curve of \cite{1989ApJ...345..245C} that 
increased the brightness in $B$, $V$, and $I_{\rm C}$ by 
4.106, 3.096, and 1.897\,mag, respectively. 

To obtain the magnitudes of the true continuum, we corrected 
them for the effect of emission lines (see Appendix~\ref{app:A}). 
For the $B$ magnitude we adopted $\Delta B = -0.3$\,mag determined 
from well-exposed spectra on days 121 and 143 that cover the whole 
passband. For $V$ we used $\Delta V$ = -0.05 and -0.1\,mag during 
the plateau phase and the maximum, respectively (green and red 
points in Fig.~\ref{fig:colors}; Table~\ref{tab:dml}). In the 
case of $I_{\rm C}$ magnitudes, we estimated 
$\Delta I_{\rm C}$ to -0.1\,mag during the maximum (day 297 and 
303), when strong \oi\,$\lambda 8446$ and $\lambda 7772$ appeared 
in the spectrum, while $\Delta I_{\rm C} \gtrsim -0.05$ was 
estimated for days 74 and 81, when these lines were absent. 
Moreover, according to SED models, we subtracted the contribution 
from the WD in the $B$ band that corresponds to $\approx$0.2\,mag 
(it is negligible in other photometric bands). 
Adding these corrections to the reddening-free $B$, $V$, and 
$I_{\rm C}$ magnitudes, we estimated color indices of the true 
continuum as ($B+0.3+0.2$) -- ($V+0.1$) = $B-V+0.4$ and 
($V+0.05/0.1$) -- ($I_{\rm C}+0.05/0.1$) = $V-I_{\rm C}$. 

Finally, we determined the theoretical color indices of the hydrogen 
nebular continuum by calculating the volume emission coefficient 
$\varepsilon_{\lambda}({\rm H},T_{\rm e})$ (see Eq.~(\ref{eq:sed})) 
at the effective wavelengths of the photometric $B$, $V$, and 
$I_{\rm C}$ passbands as a function of $T_{\rm e}$ and converted 
the corresponding fluxes to magnitudes according to the calibration 
of \cite{1982asph.book.....H} and \cite{1979PASP...91..589B}. 

A comparison of the color indices of the true continuum derived 
from the observations and the theoretical indices is shown in 
Figure~\ref{fig:colors}. 
The large scatter in the measured values ($\gtrsim 0.1$\,mag) is 
mainly due to systematic errors caused by the different instrumental 
systems used by different observers and their treatment of the 
raw data. 
Another source of uncertainty %in the magnitudes of the true continuum 
is the variability of the equivalent widths of the emission lines 
along the outburst due to variations in both the continuum and 
the lines. 

\bibliography{v1047}{}

\bibliographystyle{aasjournal}

\end{document}